\documentclass[conference]{IEEEtran}
\IEEEoverridecommandlockouts
\usepackage[utf8]{inputenc} 
\usepackage[T1]{fontenc}    
\usepackage{hyperref}       
\usepackage{url}            
\usepackage{booktabs}       
\usepackage{amsfonts}       
\usepackage{nicefrac}       
\usepackage{microtype}      
\usepackage{xcolor}         

\usepackage{graphicx}
\usepackage{textcomp}
\usepackage{xcolor}
\usepackage{amssymb}

\usepackage{booktabs} 
\usepackage{float}
\usepackage{multirow}
\usepackage{indentfirst}
\usepackage{multirow}
\usepackage{amsmath}
\usepackage{graphicx}
\usepackage{subfigure}
\usepackage[]{graphicx,indentfirst}
\usepackage{arydshln}
\usepackage{textcomp}
\usepackage{multicol}
\usepackage{subfigure}
\usepackage{bm}
\usepackage{url}
\usepackage{amsfonts}
\usepackage{mathrsfs}
\usepackage{algorithmic}
\usepackage[ruled,linesnumbered]{algorithm2e}
\newcounter{chapter}

\newtheorem{proposition}{Proposition}[chapter]
\newtheorem{definition}{Definition}[chapter]
\newtheorem{lemma}{Lemma}[chapter]
\newtheorem{proof}{Proof}[chapter]
\newtheorem{theorem}{Theorem}[chapter]
\newtheorem{corollary}{Corollary}[chapter]

\usepackage{ulem}
\usepackage{cite}
\usepackage{amsmath,amssymb,amsfonts}
\usepackage{algorithmic}
\usepackage{graphicx}
\usepackage{textcomp}
\usepackage{xcolor}
\def\BibTeX{{\rm B\kern-.05em{\sc i\kern-.025em b}\kern-.08em
    T\kern-.1667em\lower.7ex\hbox{E}\kern-.125emX}}
\begin{document}

\title{Bayesian Negative Sampling for Recommendation
}

\author{\IEEEauthorblockN{Bin Liu}
\IEEEauthorblockA{\textit{School of Electronic Information and Communications} \\
\textit{Huazhong University of Science and Technology (HUST)}\\
Wuhan, China \\
liubin0606@hust.edu.cn}
\and
\IEEEauthorblockN{Bang Wang{*}}
\IEEEauthorblockA{\textit{School of Electronic Information and Communications} \\
\textit{Huazhong University of Science and Technology (HUST)}\\
Wuhan, China \\
wangbang@hust.edu.cn}
}
\maketitle

\begin{abstract}
How to sample high quality negative instances from unlabeled data, i.e., negative sampling, is important for training implicit collaborative filtering and contrastive learning models. Although previous studies have proposed some approaches to sample informative instances, few has been done to discriminating false negative from true negative for unbiased negative sampling. On the basis of our order relation analysis of negatives' scores, we first derive the class conditional density of true negatives and that of false negatives. We next design a Bayesian classifier for negative classification, from which we define a model-agnostic posterior probability estimate of an instance being true negative as a quantitative negative signal measure. We also propose a Bayesian optimal sampling rule to sample high-quality negatives. The proposed Bayesian Negative Sampling (BNS) algorithm has a linear time complexity. Experimental studies validate the superiority of BNS over the peers in terms of better sampling quality and better recommendation performance.~\footnote{Source code and data are avaliable at \url{https://github.com/liubin06/BNS}}
\end{abstract}


\section{Introduction}
Negative sampling originates from the Positive-Unlabel (PU) problem~\cite{Jessa:2020:ML,Su:2021:IJCAI}: A training dataset, called \textit{PU-dataset}, contains both positively labelled and unlabeled instances, yet an unlabeled instance could belong to either the positive or negative class.  Negative sampling is to determine a policy for guiding how to sample an unlabeled instance from a PU-dataset, so as to effectively train downstream task models. Negative sampling can find many applications in diverse tasks, such as natural language processing (NLP)~\cite{Mikolov:2013:NIPS,Tang:2015:WWW}, computer vision (CV)~\cite{Qin:2021:AAAI,Zhao:2021:IJCAI}, as well as recommendation systems (RS)~\cite{Steffen:2014:WSDM,Zhang:2013:SIGIR,Ding:2020:NIPS}.

\par
We focus on negative sampling for recommendation. Many recommendation tasks can be formulated as how to rank unlabeled items for users, yet an unlabeled item is specific to one user, called his \textit{negative instance}. Most recommendation algorithms adopt a \textit{pairwise learning} framework to train a recommendation model with learnable users' and items' representations for the ranking computation~\cite{Steffen:2009:UAI,Wang:2019:SIGIR,Xiangnan:2020:SIGIR,Xuejiao:2020:ASC}. A pairwise comparison is to first form a training triple $(u,i,j)$ consisting of his positive instance $i$ and  negative instance $j$ for a user $u$, and the pairwise loss over all users is optimized for training a recommendation model. The dilemma of such pairwise comparisons lies in that a training triple consists of a negative instance of a user, yet such a negative instance could be potentially interested by the same user and should be recommended by the trained model. This motivates the problem of negative sampling for recommendation. that is, how to effectively sample negative instances for training a recommendation model. Many studies have shown that negative sampling is important to improving recommendation performance~\cite{Steffen:2014:WSDM,Zhang:2013:SIGIR,Ding:2020:NIPS,Park:2019:WWW,Huang:2021:KDD,Ding:2019:IJCAI,Yang:2020:KDD}.

\par
Recently, some negative sampling algorithms have been proposed for recommendation. We group them into the following categories. \textit{Static negative sampling}~\cite{Steffen:2009:UAI,Chen:2017:KDD,Mikolov:2013:NIPS,Xiangnan:2020:SIGIR,Weike:2013:IJCAI,Yu:2018:CIKM,Wang:2019:SIGIR}. Algorithms of this kind are to sample a negative instance according to a fixed sampling distribution, such as uniform sampling. \textit{Hard negative sampling}~\cite{Steffen:2014:WSDM,Zhang:2013:SIGIR,Wang:2020:WWW,Chen:2019:WWW}. Algorithms of this kind favor those negative instances with representations more similar to that of positive instances in the embedding space, for example, selecting higher scored or higher ranked instances~\cite{Steffen:2014:WSDM,Zhang:2013:SIGIR,Zhao:2015:CIKM}, however they are more likely to suffer from the negative problem, as reported in some recent studies~\cite{Ding:2020:NIPS,Qin:2021:AAAI,Zhao:2021:IJCAI}. A recent algorithm SRNS~\cite{Ding:2020:NIPS} of this kind advocates sampling a negative with a large variance of its predicted scores in the training precess. 

In this paper, we contribute to the negative sampling studies in the field of implicit CF in three aspects: (i) On the basis of order relation analysis of negatives' scores, we derive the class conditional density of true negatives and that of false negatives, and provide an affirmative answer from a Bayesian viewpoint to \textit{distinguish true negatives from false negatives} (\textbf{RQ1}). (ii) According to the  asymptotic property of the empirical distribution function, we defined a model-agnostic estimator of an instance being true negative as a \textit{quantitative negative signal measure} (\textbf{RQ2}). In particular, it is an unbiased posterior probability estimate of an instance being true negative that combines prior information (model-independent) with sample information (model-dependent). (iii) We also propose a Bayesian sampling rule to \textit{sample high-quality negative instances} (\textbf{RQ3}). It is the theoretically optimal sampling rule that minimizes the empirical sampling risk. Experiment studies validate our analysis and solution in terms of sampling quality and recommendation performance.

\section{Negative Sampling Analysis}
In this section, we use a general formulation of the personalized recommendation task to analyze the properties of negative sampling for training a recommendation model. We consider the following personalized recommendation problem, which has been intensively studied in the field~\cite{Steffen:2009:UAI,Steffen:2014:WSDM,Xiangnan:2020:SIGIR,Xiangnan:2016SIGIR}. Let $\mathcal{M}$ denote a recommendation model. Its input is an user-item interaction matrix $\mathbf{X}=[x_{ui}] \in \mathbb{R}^{M\times N}$, consisting of $M$ users and $N$ items. An element $x_{ui} = 1$ indicates a user $u$ has interacted with an item $i$; Otherwise, $x_{ui}=0$. The output is for each user his recommendation list, consisting of his un-interacted items ranked according to their predicted scores.

\par
To train the recommendation model $\mathcal{M}$, the widely used optimization objective is the following \textit{pairwise loss}:
\begin{equation}\label{Eq:PairewiseLossFunction}\
	\mathcal{L}_{loss} \equiv \max_{\Theta} \sum_{(u,i,j)} \ln \sigma(\hat{x}_{ui} - \hat{x}_{uj}) ,
\end{equation}
where for a user $u$, $\hat{x}_{ui}$ and $\hat{x}_{uj}$ is the predicted score for his already interacted item $i$ and un-interacted item $j$, respectively. $\Theta$ contains model trainable parameters, and $\lambda$ is a hyper-parameter in the learning process. In order to compute $\hat{x}_{ui}$,  some \textit{representation learning} techniques such as MF~\cite{Koren:2009:Computer} and LightGCN~\cite{Xiangnan:2020:SIGIR} can be used to learn a user representation $\mathbf{w}_u$ and an item representation $\mathbf{h}_i$, such that $\hat{x}_{ui} = \mathtt{sim}(\mathbf{w}_u, \mathbf{h}_i)$ with a similarity function $\mathtt{sim}$, e.g., a cosine or dot-product function.

\par
When training $\mathcal{M}$, negative sampling is used to construct training triples. For a user $u$, let $\mathcal{I}_u^+$ and $\mathcal{I}_u^-$ denote the set of his already interacted items, called \textit{positive instances} and the set of his un-interacted items, called \textit{negative instances}. A training triple $(u,i,j)$ is constructed as follows: For a user $u$ and one of his positive instance $i \in \mathcal{I}_u^+$, sample one of his negative instances $j \in \mathcal{I}_u^-$, viz., \textit{negative sampling}. On the one hand, although the instance $j$ is sampled from $\mathcal{I}_u^-$, it could be the case that the user $u$ actually likes it, but the un-interaction is simply due to that he had not seen it before, that is, the item $j$ is a \textit{false negative} with respect to the user $u$. On the other hand, the item $j$ is called a \textit{true negative}, if the user $u$ truly dislikes it.

\par
The \textit{stochastic gradient descent} (SGD) is often used to iteratively optimize the loss function for each training triple $(u, i, j)$. For a sampled instance $j \in \mathcal{I}_u^-$, if it is a true negative to $u$, the loss gradient with respect to its estimated score $\hat{x}_{uj}$ is computed by
\begin{equation}\label{Eq:GradientOriginal}
	\frac{\partial \mathcal{L}_{loss} }{\partial \hat{x}_{uj}} = [1-\sigma(\hat{x}_{ui} - \hat{x}_{uj})] (-1),
\end{equation}
where $\sigma(\cdot)$ is the \textsf{sigmoid} function. However, if the sampled instance $j$ is actually a false negative, we do not expect that such an incorrect sampling impacts much on the model training, especially causing an opposite gradient direction. As we have no prior knowledge about the sampled instance $j$,  we argue to replace the last term $(-1)$ in Eq.~\eqref{Eq:GradientOriginal} by a $\mathtt{sgn}(\cdot)$ function, that is,
\begin{equation}\label{Eq:Gradient}
	\frac{\partial \mathcal{L}_{loss} }{\partial \hat{x}_{uj}} = [1-\sigma(\hat{x}_{ui} - \hat{x}_{uj})] \cdot \mathtt{sgn}(j),
\end{equation}
where $\mathtt{sgn}(j)=-1$, if $j$ is a true negative to $u$, and $\mathtt{sgn}(j)=1$ is a false negative to $u$. The loss gradient Eq.~\eqref{Eq:Gradient} can be decomposed into two parts, i.e., \textit{gradient magnitude} and \textit{gradient direction}. This motivates our negative sampling analysis on what is a high quality negative: A sampled instance $j$ in a training triple $(u,i,j)$ is called a \textit{high-quality negative}, if it is both informative and unbiased.
\begin{itemize}
	\item \textbf{Informativeness}: The informativeness of a negative $j$ in a triple $(u,i,j)$ is defined as the loss gradient magnitude, i.e.,
	\begin{equation}\label{Eq:Informativeness}
		\mathtt{info}(j) =  [1-\sigma(\hat{x}_{ui} - \hat{x}_{uj})].
	\end{equation}
	\item \textbf{Unbiasedness}: The unbiasedness of a negative $j$ is defined as the probability that it is a true negative to user $u$, i.e.,
	\begin{equation}\label{Eq:Unbiasedness}
		\mathtt{unbias}(j) = \mathrm{P}(\mathtt{sgn}(j) = -1).
	\end{equation}
\end{itemize}

\par
The informativeness is directly defined as how much the negative $j$ can help updating the parameters of a recommendation model, in terms of its predicted score $\hat{x}_{uj}$. Given the predicted score $\hat{x}_{ui}$ of a positive instance, an excessively small value of $\hat{x}_{uj}$ leads to $\sigma(\hat{x}_{ui} - \hat{x}_{uj}) \rightarrow 1 $ and $\mathtt{info}(j)\rightarrow 0 $, i.e., the gradient vanishes, and little can be learned from $j$.

\par
The unbiasedness is actually defined as the probability of $j$ being true negative. We notice that the so-called uniform negative sampling~\cite{Steffen:2009:UAI} directly set $\mathtt{sgn}(j)=-1$ for all negatives, which introduces some \textit{sampling bias} in model training for $j$ being actually a false negative (positively labeled). We can understand its adverse effects in two aspects. On the one hand, a recommendation model aims at maximizing the likelihood  of pairwise comparisons between positive instances and negative instances by assigning higher scores to positives and lower scores to negatives. By assigning $\mathtt{sgn}(j)=-1$, the false negative $j$'s score will be decreased when performing stochastic gradient descent, since its gradient is directed to the negative direction. On the other hand, it treats a false negative $j$ that a user $u$ may be potentially interacted as a true negative instance, causing incorrect preference learning for a recommendation model. Therefore, how to identify true negative examples from unlabeled samples is the key research question that must be solved for negative sampling.


\section{The proposed algorithm}

\subsection{Order analysis on sampled instances} \label{Sec:Dis}
From un-interacted instances, negative sampling aims to select true negatives for model training; While the recommendation model aims to rank false negatives. Let us define \textit{negative classification} as the task of classifying an un-interacted instance as either a true negative or a false negative. Like the two sides of a coin, both negative sampling and recommendation include an implicit task of negative classification, that is, \textit{how to effectively classify a sampled negative instance?}

\par
According to the optimization objective, the implicit CF model is optimized for ranking the positive instances higher than negative instances. This suggests that the following \textit{order relation} of predicted scores might hold in general
\begin{equation}\label{Eq:OrderRelation}
	\hat{x}_{tn} \leq \hat{x}_{fn},
\end{equation}
where $\hat{x}_{tn}$ and $\hat{x}_{fn}$ is the predicted score of a true negative and a false negative (i.e., true positives in future testing data), respectively. Note that this is also the optimization objective for contrastive-based learning methods~\cite{gutmann:2012:JMLR,Oord:2018:arxiv,Gutmann:2010:ICAIS} that essentially encourage learned feature representation
for positive instance to be similar with "anchor" data point , while pushing features from the randomly sampled negative instance apart from "anchor" data point~\footnote{~The "anchor" here refers to the user embedding.} in the embedding space~\cite{Wang:2020:ICML,Xu:2022:arxiv,liu2021self}.

\subsection{Distribution Analysis}
Since the scores of true negatives and false negative are predicted using the same score function in a recommendation model, it is safe to assume that $\hat{x}_{fn}$ and $\hat{x}_{tn}$ are identically and independently distributed with a same probability density function $f(x)$ and a same cumulative distribution function $F(x) = P(X \leq x)$.
%

\par
Consider two IID random variables $X_{tn}$ and $X_{fn}$ with their corresponding realizations $\hat{x}_{tn}$ and $\hat{x}_{fn}$  that are sorted in an ascending order:
\begin{equation}\label{Eq:OrderStatistics}
	X_{tn} \le X_{fn}.
\end{equation}
For the two random variables $X_{tn}$ and $X_{fn}$, there exists a sufficient small interval $dx$, where one and only one random variable $X_{tn}$ with its realization $\hat{x}_{tn} \in [\hat{x}_{tn}, \hat{x}_{tn}+dx]$. Since $\hat{x}_{tn} \leq \hat{x}_{fn}$, the random variables $X_{fn}$ with its realization $\hat{x}_{fn} \in (\hat{x}_{tn}+dx,\infty )$. The probability differential of $X_{tn}$ can be computed by
\begin{equation}\label{Eq:TNdiff}
	\begin{aligned}
		g(\hat{x}_{tn})dx = &2!\times P(\hat{x}_{tn} \leq X_{tn}\leq \hat{x}_{tn} +dx)\\
	&	\times P(\hat{x}_{tn}+dx \leq X_{fn} \le \infty ) + o(dx) \\
		= &2 f(\hat{x}_{tn})dx [1 - F(\hat{x}_{tn}+dx)] + o(dx),
	\end{aligned}
\end{equation}
where $g(\hat{x}_{tn})$ is the class conditional density of true negatives. $f(\hat{x}_{tn})dx$ evaluates the probability of  $X_{tn} \in [\hat{x}_{tn}, \hat{x}_{tn}+dx]$, and $[1 - F(\hat{x}_{tn}+dx)]$ evaluates the probability of the rest  random variables $X_{fn} \in (\hat{x}_{tn}+dx,\infty)$. $o(d{x})$ is the high-order infinitesimal of $d{x}$. Dividing both sides of the equation by $dx$, the class conditional density of true negatives can be written as:
\begin{equation}\label{Eq:TNpdf}
	\begin{aligned}
		g(\hat{x}_{tn}) =& \lim_{dx \to 0} \frac{2 f(\hat{x}_{tn})dx [1 - F(\hat{x}_{tn}+dx)] + o(dx)}{dx} \\
		=&2 f(\hat{x}_{tn}) [1 - F(\hat{x}_{tn})].		
	\end{aligned}
\end{equation}

Likewise, the distribution of false negatives is given by:
\begin{equation}\label{Eq:FNpdf}
	\begin{aligned}
		h(\hat{x}_{fn})
		= 2 F(\hat{x}_{fn}) f(\hat{x}_{fn}).
	\end{aligned}
\end{equation}

\begin{proposition} \label{orderdist}
	If $f(x)$ is a probability density function, $F(x) = \int_{-\infty}^{x} f(t)dt$ is the corresponding cumulative distribution function, then \\
	(i) $g(x)=2 f(x)[1-F(x)]$ is a probability density function that satisfies $g(x) \geq 0$ and $\int_{-\infty}^{\infty}  g(x)dx =1$. \\
	(ii) $h(x)=2 f(x) F(x)$ is a probability density function that satisfies $h(x) \geq 0$ and $\int_{-\infty}^{\infty}  h(x)dx =1$.
	\begin{proof}
		Since $f(x) \geq 0$, $F(x)= \int_{-\infty}^{x} f(t)dt \in [0,1]$, so $g(x) = 2f(x) [1-F(x)] \geq  0$, $h(x) =2f(x)F(x) \geq 0$.
		\begin{eqnarray}
			\int_{-\infty}^{\infty}  g(x)dx &=& \int_{-\infty}^{\infty} 2 f(x)[1-F(x)] dx \nonumber \\
			&=& 2 \int_{-\infty}^{\infty}  f(x)dx - 2\int_{-\infty}^{\infty}  f(x) F(x)dx \nonumber\\
			&=&2 -  2 \int_{-\infty}^{\infty}   F(x)dF(x) \nonumber \\
			&=&1, \nonumber \\
			\int_{-\infty}^{\infty}  h(x)dx &=& \int_{-\infty}^{\infty} 2 f(x) F(x) dx \nonumber \\
			&=&  2\int_{-\infty}^{\infty} F(x)dF(x) \nonumber\\
			&=& 1. \nonumber
		\end{eqnarray}	
	\end{proof}
\end{proposition}

\subsubsection{Real Distribution}\label{Appendix:Realdist}
\begin{figure*}[h!]
	\centering
	\includegraphics[width=\textwidth]{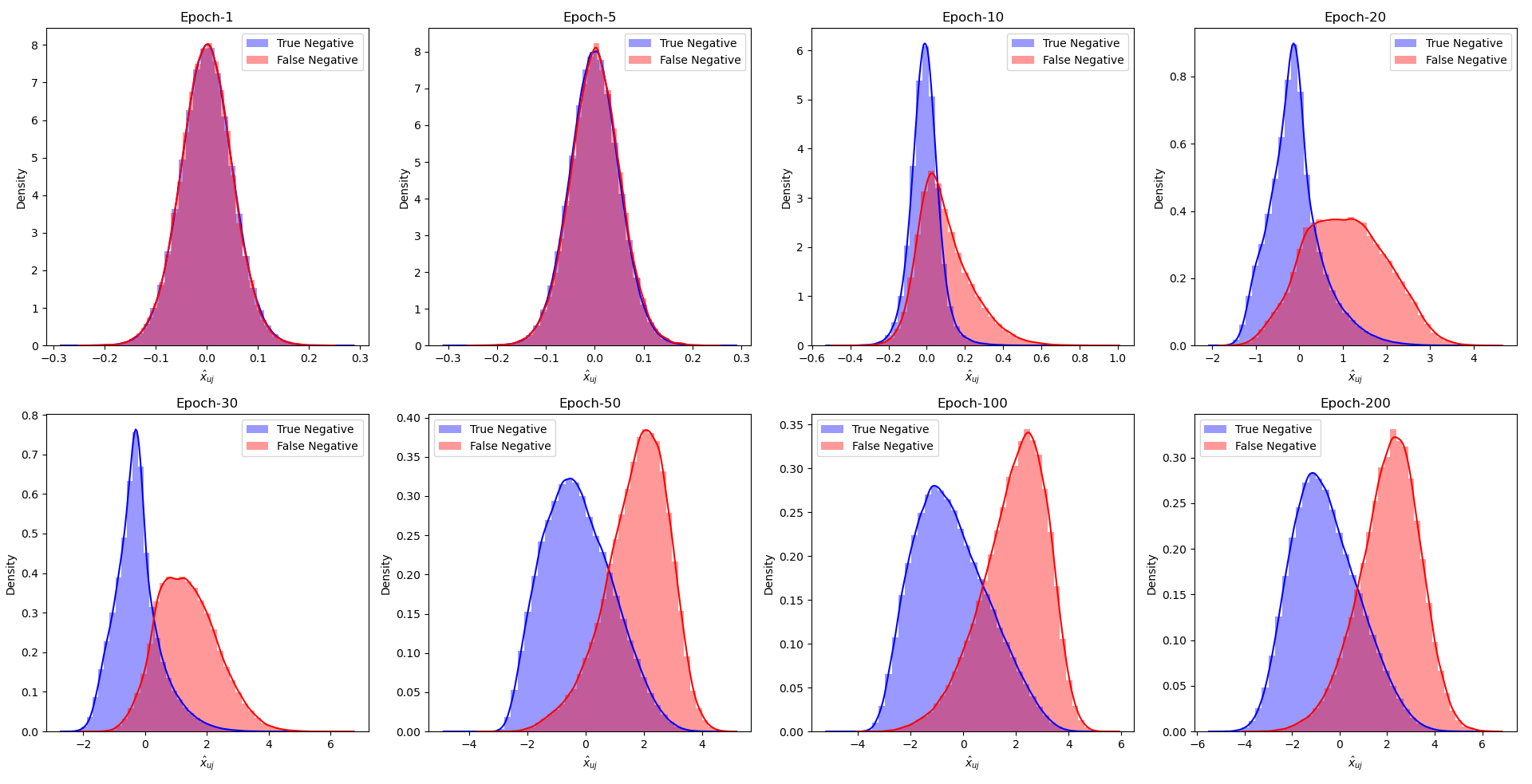}
	\caption{Real distribution of TN and FN at different training epochs on the MovieLens-100K dataset by using the uniform random negative sampling.}
	\label{Fig:DistributionStatistics}
\end{figure*}

To verify the order relation of Eq~\eqref{Eq:OrderRelation}, we use the ground-truth labels for instances in the test set to obtain the false negatives that are positively labeled but unobserved during the training process. And the rest of un-interacted items are true negatives. We adopt the classic \textit{matrix factorization} recommendation model with random negative sampling to train the model on the training set, and record their predicted scores at each training epoch. By counting the predicted scores of true negatives and false negatives, we report their distribution densities in the model training by Fig.\ref{Fig:DistributionStatistics}.

\par
Fig.~\ref{Fig:DistributionStatistics} provides two insightful findings:
(a) The higher the predicted score of a negative instance, the higher the probability density that it is a false negative and the lower the probability density that it is a true negative;
(b) As the training continues, the distinction between two distributions gradually becomes clearer: Compared to that of true negatives, the distribution of false negatives is centered on a larger score. This suggests that the score function of a recommendation model is capable of rating the false negatives higher than true negatives.

\subsubsection{Theoretical Distribution}\label{Appendix:Theodist}
\begin{figure}[h!]
	\centering
	\includegraphics[width=8.5cm]{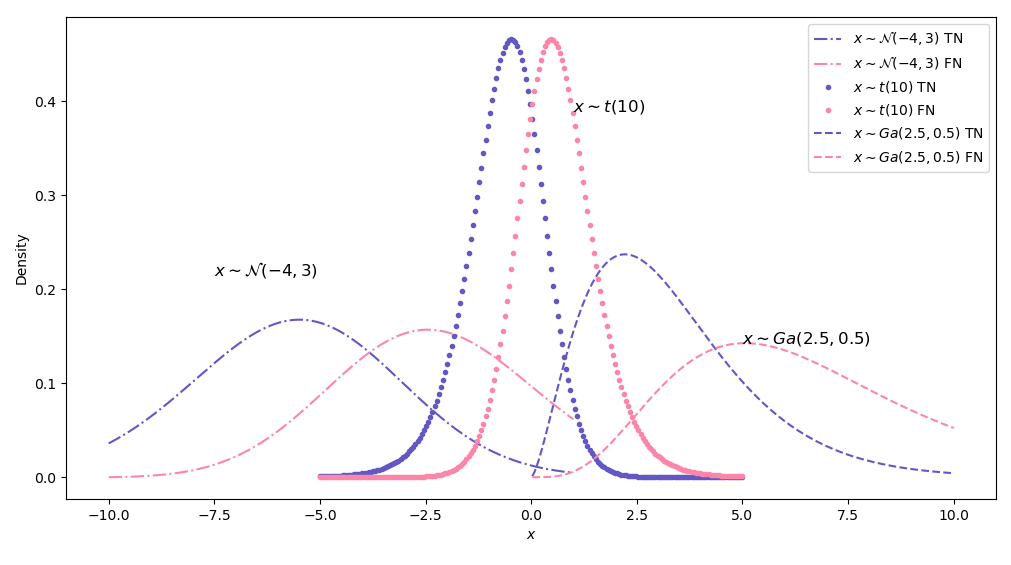}
	\caption{Theoretical distribution of TN and FN with different types of $f(x)$.}
	\label{Fig:TheoryDist}
\end{figure}
Fig~\ref{Fig:TheoryDist} exhibits the distribution morphology of false negatives and true negatives derived from the ordinal relation with different types of $f(x)$: Gaussian distribution $x\sim \mathcal{N}(\mu,\sigma)$ (symmetrical), student distribution $x \sim t(n)$ (symmetrical), and Gamma distribution $x\sim Ga(\alpha,\lambda)$ (asymmetrical) . As we can see, during the training process, the actual distribution of true/false negatives in Fig~\ref{Fig:DistributionStatistics}  gradually exhibit the same structure as depicted in Fig~\ref{Fig:TheoryDist}. Different scoring functions and ranking models yield different density expressions $f(\cdot)$, but this separated structure is sufficient for us to classify true negatives and false negatives.

So far, we do not know the explicit expression for the $f(x)$ and $F(x)$. Yet the calculation of empirical distribution function  $ F_{n}(x)= \frac{1}{n}\sum_j I_{|X_\cdot \leq \hat{x}_l|}$ is easy to implement, which  converges to common cumulative distribution function $F(x)$ almost surely by the strong law of large numbers. Glivenko theorem~\cite{glivenko:1933} strengthened this result by proving uniform convergence of $F_n(\cdot)$ to $F(\cdot)$. This conclusion helps us for calculating abstract function $F(\cdot)$.


\subsection{Bayesian Negative Classification}
For an un-interacted item $l$ with its predicted score $\hat{x}_l$, the posterior probability of the item $l$ being a true negative can be computed using the Bayesian formula:
\begin{eqnarray} \label{Eq:PostTN}
	P(tn|\hat{x}_l) &\propto& P(\hat{x}_l|tn) P_{tn}(l) \nonumber \\
	&=&  2f(\hat{x}_l) [1 - F(\hat{x}_l)] P_{tn}(l),
\end{eqnarray}
where $P(\hat{x}_l|tn)$ is the class conditional density of true negatives given by $g(\hat{x})$,  $P_{tn}(l) = 1- P_{fn}(l) $ is the prior probability of item $l$ being true negative instance. $f(\hat x_l)$ is the score distribution of un-interacted items, $F(\hat x_l) = \int_{-\infty}^{\hat{x}_l} f(t) dt$ is the corresponding cumulative distribution function.

\par
Also, the posterior probability of item $l$ being a false negative can be computed by:
\begin{eqnarray}\label{Eq:PostFN}
	P(fn|\hat{x}_l) &\propto& P(\hat{x}_l|fn) P_{fn}(l) \nonumber \\
	&=& 2 F(\hat{x}_l) f(\hat{x}_l) P_{fn}(l)
\end{eqnarray}

\par
Generally, the Bayesian classifier can be obtained by maximizing the posterior probability:
\begin{eqnarray}
	\mathop{\arg\max}\limits_{c \in \{fn, tn\}} P(c|\hat{x}_l).
\end{eqnarray}

Directly computing the score density function $f(\cdot)$ in Eq.~\eqref{Eq:PostTN} and Eq.~\eqref{Eq:PostFN} is complicated. Yet the calculation of empirical distribution function  $ F_{n}(x)= \frac{1}{n}\sum_j I_{|X_\cdot \leq \hat{x}_l|}$ is easy to implement. So we define unbiasedness of an un-interacted item $l$ in a fractional form to eliminate the density function $f(\cdot)$:
\begin{eqnarray}
	\mathtt{unbias}(l) &=& \frac{P(tn|\hat{x}_l)}{P(tn|\hat{x}_l)+P(fn|\hat{x}_l)} \label{Eq:NorPost}  \\
	&\propto& \frac{ f(\hat{x}_l) [1 - F(\hat{x}_l)] P_{tn}(l)}{ f(\hat{x}_l) [1 - F(\hat{x}_l)] P_{tn}(l) + F(\hat{x}_l) f(\hat{x}_l) P_{fn}(l) }  \nonumber\\
	&=&  \frac{  [1 - F(\hat{x}_l)][1-P_{fn}(l)] }{1 - F(\hat{x}_l) -P_{fn}(l) + 2F(\hat{x}_l)P_{fn}(l) }.\label{Eq:unbias}
\end{eqnarray}
According to Glivenko theorem~\cite{glivenko:1933}, $F(\hat{x})$ can be approximated using $F_n(\hat{x}_l)$, i.e., the percentage of $\hat{x}_{\cdot} \leq \hat{x}_l$ 
\begin{eqnarray}\label{Eq:CDF}
	F(\hat{x}_l) = \frac{ \# \{\hat{x}_{ \cdot} \leq \hat{x}_l |~ l \in \mathcal{I}_u^- \} } {\# \{\mathcal{I}_u^- \} }.
\end{eqnarray}
$P_{fn}(l)$ is the prior probability of $l$ being false negative. For convenience, we assume the times of item $l$ being interacted $pop_l \sim B (N, P_{fn}(l))$, where $N$ is  
total number of interactions in training set. So
\begin{eqnarray}\label{Eq:Prior}	
	P_{fn}(l) = \frac{pop_l}{N}.
\end{eqnarray}

\begin{lemma}[Unbiased negative signal]\label{unbias}
	If $pop_l \sim B (N, P_{fn}(l))$, then $\mathtt{unbias}(l)$ measure given by Eq~\eqref{Eq:unbias} is an unbiased estimator for $l$ being true negative.
	
	\begin{proof}
	Setting random variable $Y=1$ if $l \in fn$, otherwise $Y=0$,
	\begin{eqnarray}
		Y= \left\{
		\begin{aligned}
			1 ,~ P &=& \theta \\
			0 ,~ P &=& 1-\theta, \\
		\end{aligned}
		\right.
	\end{eqnarray}
	where $\theta$ is the probability of $l$ being false negative. So $Y \sim B(1,\theta)$. $pop_l = Y_1 + Y_2 + \cdots + Y_N  \sim B(N,\theta)$, then $P(pop_l=k )= \binom{N}{k} \theta^k (1-\theta)^{(N-k)} $. So
	
	\begin{eqnarray}
		\mathbb{E}[P_{fn}(l)] &=& \mathbb{E}(\frac{pop_l}{N}) \nonumber \\
		&=& \theta
	\end{eqnarray}
	
	Given the observation $\hat{x}_{ul} = X$, $F(X)$ is a statistic of $X$. $P_{fn}(l)$ is a statistic of $\sum_i Y_i$ that is independent of $X$. So
	\begin{eqnarray}
		&&\mathbb{E} [\mathtt{unbias}(l)]  \nonumber \\
		&=& \mathbb{E}   \frac{  [1 - F(X)][1-P_{fn}(l)] }{1 - F(X) -P_{fn}(l) + 2F(X)P_{fn}(l) } \nonumber  \\
		&=&  \frac{  [1 - \mathbb{E}  [F(X)]] [1- \mathbb{E} [ P_{fn}(l)]] }{1 -\mathbb{E} [F(X)] - \mathbb{E}[P_{fn}(l)] + \mathbb{E} [2F(X)P_{fn}(l)] } 
	\end{eqnarray}
	$\mathbb{E}[F(X)]$ is the first order origin moment of cumulative distribution function $F(X)$
	\begin{eqnarray}
		\mathbb{E}  [F(X)] &=& \int_{-\infty}^{\infty} F(x) f(x) dx\nonumber\\
		&= &  \int_{-\infty}^{\infty} F(x) dF(x)\nonumber\\
		&= &  \frac{1}{2}  F^2(x)   \bigg|_{x=-\infty}  ^{x=\infty} \nonumber\\
		&= & \frac{1}{2}.
	\end{eqnarray}
	So 
	\begin{eqnarray}
		\mathbb{E} [\mathtt{unbias}(l)] &=&  \frac{  (1 - \frac{1}{2}) (1-\theta) }{1 -\frac{1}{2} - \theta + 2\cdot \frac{1}{2} \cdot \theta} \nonumber  \\
		&=& 1-\theta.
	\end{eqnarray}
	Note $1-\theta$ is the probability of $Y=0$ from binomial populations $Y\sim B(1,\theta)$, so $\mathtt{unbias}(l)$ is unbiased estimator of $l$ being true negative. Fig~\ref{Fig:unbias} plots $\mathtt{unbias}(l)$ as a function of $F(\hat{x}) \in [0,1]$ and $P_{fn}(l) \in [0,1]$. We observe that $\mathtt{unbias}(l)$ is a decreasing function w.r.t both $F(\hat{x})$ and $P(fn)$. The monotonicity of $\mathtt{unbias}(j)$ is consistent with our analysis, and the value domain of $\mathtt{unbias}(j)$ $\in [0,1]$ conforms to the probability form.  
	\begin{figure}[!]
		\centering
		\includegraphics[width=8.8cm]{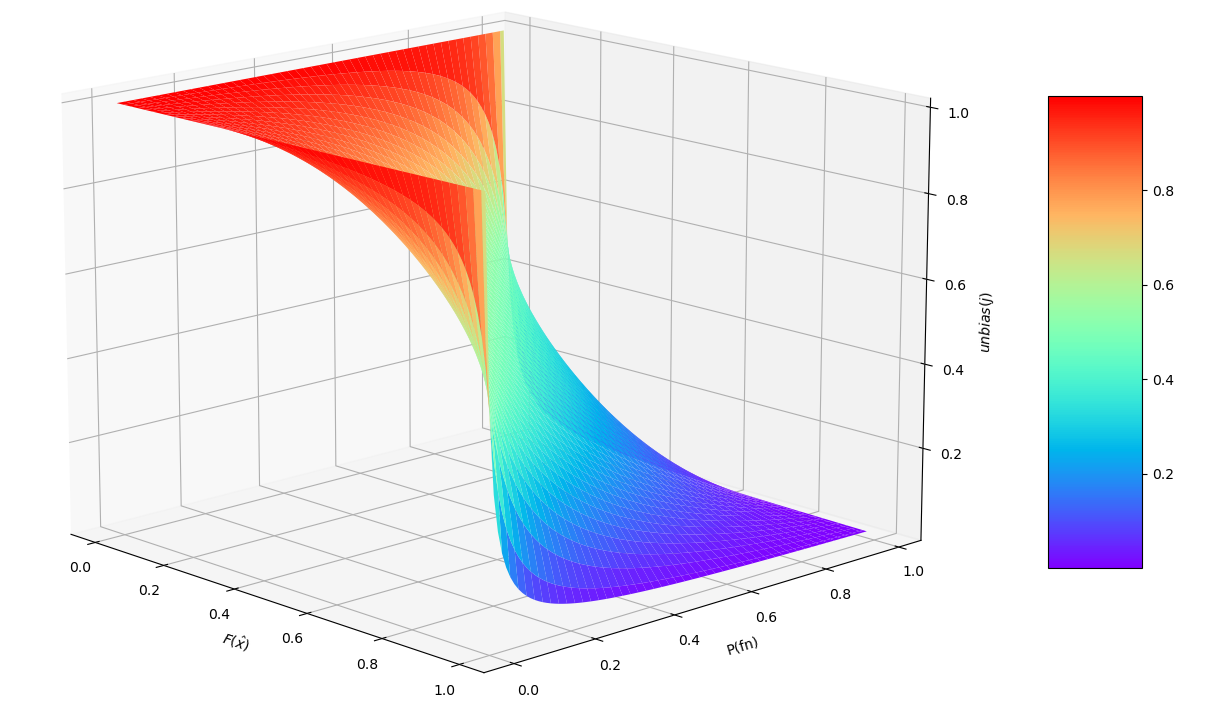}
		\caption{Numerical plots of normalized posterior probability $\mathtt{unbias}(j)$ by Eq.~\eqref{Eq:unbias}.}
		\label{Fig:unbias}
	\end{figure}
	\end{proof}
\end{lemma}

The definition of unbiasedness can be interpreted as the \textit{normalized posterior probability} of $l$ being true negative (negative signal). The density function $f(\hat{x}_l)$ is eliminated due to the fractional expression of Eq.~\eqref{Eq:unbias}. This suggests that it is model-independent metric and can be generalized to different ranking models. Eq.~\eqref{Eq:unbias} implies that the negative signal of an un-interacted item $l$ is formally determined by: (a) Non-threshold classification result of a ranking model denoted by $F(\hat{x}_l)$~\footnote{ Ranking position and $F(\hat{x}_l)$ are with a one-to-one mapping.}. A larger $F(\hat{x})$, i.e., a higher ranking position, indicates that to a large extent a ranking model classifies $l$ as a false negative. Now we provide a new understanding of $F(\hat{x}_l)$. $F(\hat{x}_l)$ describes the joint probability of the observed sample  $\hat{x}_l$ as a function of the parameters of the ranking model. For the specific parameter $l \in fn$, $F(\hat{x}_l)$ assigns a probabilistic prediction valued in $[0,1]$ of $l$ being false negative (positive).  This explains why those hard negative sampling algorithms~\cite{Steffen:2014:WSDM,Zhang:2013:SIGIR} for over-sampling higher scored or higher ranked items are more likely to suffer from the false negative problem. (b) Prior information denoted by prior probability ($P_{tn}(l)$ or $P_{fn}(l)$).

%

We note that the metric of Eq.~\eqref{Eq:unbias} covers two major paradigms of current works on distilling negative signals: (i) modeling prior information using exposure data~\cite{Jingtao:2019:IJCAI}, KG entities~\cite{Wang:2020:WWW}, connections in social networks~\cite{Zhao:2014:CIKM,Wang:2016:CIKM}, etc. (ii) utilizing the sample information from ranking model such as predicted score~\cite{Steffen:2014:WSDM}, ranking position~\cite{Zhang:2013:SIGIR}, scores' variance~\cite{Ding:2020:NIPS}, etc. The former can incorporate domain knowledge, however the negative signal is independent of model status; The latter exploits sample information $\hat{x}_l$ but ignores the prior information. While being easy to sample informative instances, methods of this kind are likely to suffer false negative problems. The advantage of $\mathtt{unbias}(\cdot)$ measure lies in the theoretical foundation of posterior probability, which combines priori information (model-independent) and sample information $\hat{x}_l$ (model-dependent). For the sake of simplicity and completeness of the theoretical proof, we adopted a naive approach for modeling prior probability. Some other additional information and domain knowledge can also be exploited for modeling $P_{tn}(l)$. 

\subsection{Negative Sampling Algorithm}
The ranking objective of Eq~\eqref{Eq:PairewiseLossFunction} is the analogy to $AUC$ metric~\cite{Steffen:2009:UAI}. It replaces the non-differentiable Heaviside function used in $AUC$ metric with the differentiable loss $\ln \sigma(\cdot)$, which is a common practice when optimizing for $AUC$~\cite{Herschtal:2004:ICML,Steffen:2009:UAI}. When performing a single negative sampling, the unlabeled instance $l$ is directly assigned a negative gradient, resulting in a \textit{minus effect} of the predicted score $\hat{x}_{ul}$, which will have two effects on $\mathcal{L}$: (i) decreasing of $\mathcal{L}$ by sampling a false negative, denoted as $\triangle \mathcal{L}_{fn}(l|i)$; (ii) increasing of $ \mathcal{L}$ by sampling a true negative, denoted as $\triangle \mathcal{L}_{tn}(l|i) $.

\begin{definition}[Conditional sampling risk]
	Given the positive instance $i$, we define the conditional sampling risk for sampling  $l$ as the expectation of sampling loss $\triangle \mathcal{L}(l|i)$ over the distribution of $l$'s ground truth label $c$:
	\begin{eqnarray}
		&&R(l|i) \nonumber \\
		 &=&  \mathbb{E}_{l \sim P(c|l)}  \triangle \mathcal{L}(l|i)\nonumber\\ 
		&=&[1- P(tn|l)] \cdot \triangle \mathcal{L}_{fn}(l|i) + P(tn|l)\cdot \triangle \mathcal{L}_{tn}(l|i),
	\end{eqnarray}
	where $P(tn|l)$ is the posterior probability of $l\in tn$ and can be computed by $\mathtt{unbias}(l)$, and $ 1 - P(tn|l)$ is the posterior probability of $l\in fn$. 
\end{definition}

\begin{definition}[Empirical sampling risk]
	Given a set of positive instances, we define the empirical sampling risk for sampler $h $  as the expectation of $R(l|i)$ over the distribution of positive instances:
	\begin{eqnarray}
		R(h) = \mathbb{E}_{i \sim P(i)} R(l|i).
	\end{eqnarray}
\end{definition}

\begin{theorem}[Optimal sampling rule] \label{optimalrule}
	If the conditional sampling risk $R(l|i)$ is independent of each other, for any sampler $h: \mathcal{I}_u^- \rightarrow l$,
	\begin{eqnarray}\label{Eq:OptimalSam}
		h^* &=&   \mathop{\arg\min}\limits_{l \in\mathcal{I}_u^-} R(l|i) 
	\end{eqnarray}
	uniformly superior than $h$ that minimizes the empirical sampling risk.
	\begin{proof}
		Given the training set, the distribution of positive samples $P(i)$ is determined. The empirical sampling risk is 
		\begin{eqnarray}
			R(h) = \mathbb{E}_{i \sim P(i) } R(h|i)
		\end{eqnarray}
		where $R(h|i)$ is the conditional sampling risk given positive instance $i$. Then
		\begin{eqnarray}
			&&	R(h^*) - R(h) \nonumber\\
			&=&	\mathbb{E}_{i \sim P(i) }  R(h^*|i) - \mathbb{E}_{i \sim P(i) }  R(h|i)  \nonumber \\
			&=& \sum_i P(i)  [R(h^*|i) - R(h|i)] \nonumber \\
			&=& \sum_i P(i)  [  \mathop{\arg\min}\limits_{l \in\mathcal{I}_u^-} R(l|i)  - R(h|i)]\nonumber \\
			&\leq& 0.
		\end{eqnarray}
		So the infimum of empirical sampling risk can be given in the form of optimal sampler $h^*$:
		\begin{eqnarray}
			\inf \{R(h)\} 	&=& R(h^*) \nonumber\\
			&=& \mathbb{E}_{i \sim P(i) } R(h^*|i).
		\end{eqnarray}
	\end{proof}
\end{theorem}
The conclusion is concise and explicit: if the sampler $h^*$ minimizes the conditional sampling risk $R(l|i)$, then the empirical sampling risk will be minimized. In turn, we can measure the quality of negative examples in terms of conditional sampling risk.

\begin{corollary}\label{highquality}
	For two negative instances $l,l^{'} \in \mathcal{I}_u^-$ ,  $l$ is superior than $l^{'}$ if $R(l|i) \leq R(l^{'}|i)$ .
\end{corollary}

Next we will estimate sampling loss $\triangle \mathcal{L}(l|i)$ for sampling negative instance $l$. To simplify the analysis, we follow~\cite{Steffen:2009:UAI} to adopt the independence assumption, and consider the \textit{minus effect} on a single pair's ranking objective $\tilde{\mathcal{L}}= \ln \sigma( \hat{x}_{fn} - \hat{x}_{tn})$. The Taylor expansion of ranking objective of $\tilde{\mathcal{L}}$ around point $\hat{x}_{ul}$ is

\begin{eqnarray}
	\tilde{	\mathcal{L}}' =\left\{
	\begin{aligned}
		\tilde{\mathcal{L}} +  \frac{\partial \mathcal{L}} {\partial \hat{x}_{ul}}  ( \hat{x}_{ul}' - \hat{x}_{ul}) + o( \hat{x}_{ul}' - \hat{x}_{ul}) ,~ &if&~ l = fn\\
		\tilde{\mathcal{L}} -  \frac{\partial \mathcal{L}} {\partial \hat{x}_{ul}}  ( \hat{x}_{ul}' - \hat{x}_{ul}) + o( \hat{x}_{ul}' - \hat{x}_{ul}) ,~ &if&~ l = tn.\\
	\end{aligned}
	\right.
\end{eqnarray} 
where $ \frac{\partial \mathcal{L}} {\partial \hat{x}_{ul}} =  \mathtt{info}(l) $. So the unit decrease of $\hat{x}_{ul}$ (i.e., $\triangle \hat{x}_{ul}=-1$) results in $\triangle \mathcal{L}_{fn}(l|i) = \tilde{\mathcal{L}}-\tilde{\mathcal{L}}' \approx \mathtt{info}(l)$, indicating the decrease of $\tilde{\mathcal{L}}$ if $l$ is false negative (positively labeled); Otherwise $\mathcal{L}_{tn}(l|i)  \approx -\mathtt{info}(l) $, indicating the increase of $\tilde{\mathcal{L}}$. To take the overall ranking list into consideration, we introduce a hyperparameter $\lambda$ to control the effect scale, and estimate the sampling loss as
\begin{eqnarray} \label{Eq:rankinggain}
	\triangle	\mathcal{L}(l|i)  \approx \left\{
	\begin{aligned}
		\mathtt{info}(l) ,~ &if&~ l = fn\\
		- \lambda \mathtt{info}(l) ,~ &if&~ l = tn\\
	\end{aligned}
	\right.
\end{eqnarray}

So the conditional sampling risk for sampling instance $l$ given positive instance $i$ is
\begin{eqnarray}
	R(l|i) = P(fn|l) \cdot \mathtt{info}(l) - P(tn|l)\cdot \lambda\mathtt{info}(l).
\end{eqnarray}
Based on Corollary~\ref{highquality}, we select high quality negative instance $j$ by:
\begin{eqnarray} \label{Eq:NegativeSam}
	j   &=&   \mathop{\arg\min}\limits_{l \in \mathcal{M}_u} R(l|i) \nonumber \\
	&=& \mathop{\arg\min}\limits_{l \in \mathcal{M}_u}~ [1-\mathtt{unbias}(l)] \cdot \mathtt{info}(l) \nonumber\\  
	&&- \lambda \cdot \mathtt{unbias}(l) \cdot \mathtt{info}(l)  \nonumber \\
	&=& \mathop{\arg\min}\limits_{l \in\mathcal{M}_u}~ \mathtt{info}(l)\cdot [1-(1+\lambda)\mathtt{unbias}(l)]
\end{eqnarray}
where $\mathcal{M}_u \subseteq  \mathcal{I}_u^-$ is a small candidate set containing randomly selected  negative instances from $\mathcal{I}_u^-$. When $|\mathcal{M}_u| = |\mathcal{I}_u^-|$, $h = h^*$; When $\lambda \rightarrow \infty$, $h$ reduces to $\mathop{\arg\max}\limits_{l \in\mathcal{M}_u} \mathtt{info}(l)\cdot \mathtt{unbias}(l)$, i.e., sampling those both informative (\textit{hard}~\footnote{Similar to positive instances in the embedding space.}) and unbiased (\textit{negative}) instances.


\par
\textbf{Complexity}: We summarize the proposed negative sampling algorithm as: for each negative instances in the candidate set $\mathcal{M}_u$, (i) computing interaction ratio as prior probability ($\mathcal{O}(1)$), (ii) computing $F(\hat{x}_l)$ as probabilistic prediction by ranking model (likelihood) ($\mathcal{O}(|\mathcal{I}|)$), (iii) computing the negative signal $\mathtt{unbias}(l)$ as posterior probability ($\mathcal{O}(1)$), and (iv) perform Bayesian negative sampling based on Eq~\eqref{Eq:NegativeSam} ($\mathcal{O}(1)$). So the proposed \textsf{BNS} has a linear time complexity. Algorithm~\ref{Alg:1}  provides the pseudo-codes of our proposed negative sampling algorithm. 

\begin{algorithm}[h]\label{Alg:1}
	\normalem
	\caption{The proposed Bayesian negative sampling (BNS) algorithm}
	\label{Alg}
	\KwIn{Traning set $\mathcal{R}=\{(u,i)\}$, score function $s(\cdot)$, $m$ (size of $\mathcal{M}_u$ ), embedding size $d$, weight $\lambda$.}
	\KwOut{User embedding $\{\mathbf{w}_u | u \in \mathcal{U}\}\in \mathbb{R}^d$, item embedding $\{\mathbf{h}_i | i \in \mathcal{I}\}\in \mathbb{R}^d$}
	\For{$epoch=1, 2, ..., T$}{
		Sample a mini-batch $\mathcal{R}_{batch} \in \mathcal{R}$\\
		\For{each $(u,i) \in \mathcal{R}_{batch}$}{
			Get rating vector $\hat{\mathbf{x}}_u$ . \label{Algo:rating}\\
			$\backslash$$\backslash$ $\textbf{Starting~Negative~Sampling}$ \\
			Uniformly sample candidate set $\mathcal{M}_u \subseteq  \mathcal{I}_u^-$. \label{Algo:candi} \\
			\For {each $(u,l) \in \mathcal{M}_u$}{
				Calculate $\mathtt{info}(l)$ by Eq~\eqref{Eq:Informativeness}. \label{Algo:inf} \\
				Calculate $P_{fn}(l)$ by Eq~\eqref{Eq:Prior}. $\backslash$$\backslash$~\textit{prior} \label{Algo:p} \\
				Calculate $F(\hat{x}_l)$ by Eq~\eqref{Eq:CDF}. $\backslash$$\backslash$~\textit{likelihood}\label{Algo:f} \\	
				Calculate $\mathtt{unbias}(l)$ by Eq~\eqref{Eq:unbias}. $\backslash$$\backslash$~\textit{posterior} \label{Algo:unbias} \\
			}
			Sampling negative $j$ based on strategy Eq~\eqref{Eq:NegativeSam}. \label{Algo:samp} \\
			Update embeddings $\mathbf{w}_u, \mathbf{h}_i, \mathbf{h}_j$.}
	}
	\KwResult{Final embeddings.}
\end{algorithm}

%

\section{Experiment}
\subsection{Experiment Settings}
\subsubsection{Dataset}
We conduct experiments on three public datasets, including MovieLens-100k\footnote{~\url{https://grouplens.org/datasets/movielens}}, MovieLens-1M, and Yahoo!-R3.\footnote{~\url{http://webscope.sandbox.yahoo.com/catalog.php?datatype=r}.}~\cite{Xuejiao:2020:ASC} They contain users' ratings on items according to a discrete five-point grading system on an interacted item. In the research of personalized ranking from implicit feedbacks~\cite{Steffen:2009:UAI,Yu:2018:CIKM,Weike:2013:IJCAI,Xuejiao:2020:ASC,Zhang:2013:SIGIR,Steffen:2014:WSDM}, a commonly used technique for data preprocessing is to convert all rated items to implicit feedbacks. Following their approaches, our training sets only consist of interacted items but not with their rating details. For each dataset, we randomly select 20\% as test data, and the rest 80\% as training data. Table~\ref{Table:Dataset} summarizes the dataset statistics.

\subsubsection{Baselines}
We compare the proposed algorithm with three types of negative sampling methods: (a) Fixed negative sampling distribution, including \textsf{RNS}\cite{Steffen:2009:UAI,Xiangnan:2020:SIGIR,Weike:2013:IJCAI,Yu:2018:CIKM,Wang:2019:SIGIR,Xuejiao:2020:ASC,Liu:2017:SIGIR}: and \textsf{PNS}\cite{Mikolov:2013:NIPS,Chen:2017:KDD,Tang:2015:WWW}, (b) hard negative sampling with dynamic sampling distribution, including \textsf{AOBPR}\cite{Steffen:2014:WSDM} and \textsf{DNS}\cite{Zhang:2013:SIGIR}, and hard negative sampling based on priori statistical information to oversample high-variance negatives, including \textsf{SRNS}\cite{Ding:2020:NIPS}. We note that all these competitors also only use the positive and unlabeled data but without additional information for negative sampling. 
\begin{itemize}
	\item[-]\textsf{RNS}\cite{Steffen:2009:UAI,Xiangnan:2020:SIGIR,Weike:2013:IJCAI,Yu:2018:CIKM,Wang:2019:SIGIR,Xuejiao:2020:ASC,Liu:2017:SIGIR}: (Random Negative Sampling) Uniformly sampling negatives.
	\item[-]\textsf{PNS}\cite{Mikolov:2013:NIPS,Chen:2017:KDD,Tang:2015:WWW}: (Popularity-biased Negative Sampling) Adopting a fixed distribution proportional to an item interaction ratio, i.e., $\propto r_j ^{.75}$.
	\item[-]\textsf{AOBPR}\cite{Steffen:2014:WSDM}: Over-sampling global higher ranked  negatives with sampling probability proportional to $ exp(-rank(j|u)/\lambda)$, where $rank(j|u)$ is the ranking position of predicted score $\hat{x}_{uj}$ in $u$'s predicted score vector $\hat{\mathbf{x}}_u$.
	\item[-]\textsf{DNS}\cite{Zhang:2013:SIGIR}: Over-sampling relative higher ranked hard negatives. Its sampling probability is a liner function to the relative ranking position.
	\item[-]\textsf{SRNS}\cite{Ding:2020:NIPS}: Using priori statistical information for over-sampling high-variance negatives.
\end{itemize}

\begin{table}[h!]
	\centering
	\caption{Dataset Statistics}\label{Table:Dataset}
	\begin{tabular}{lrrrr}
		\toprule[1.2pt]
		~           & users   & items   & training set  &test set  \\ \cline{1-5}
		MovieLens-100k   &   943    &  1,682   &    80k	   & 20k 	\\
		MovieLens-1M    &   6,040  &  3,952   &   800k     & 200k  \\
		Yahoo!-R3       &   5,400  &  1,000   &   146k      & 36k  \\
		\bottomrule[1.2pt]
	\end{tabular}
\end{table}

\subsubsection{Experimental setup}
We use the classic \textit{matrix factorization} (MF)~\cite{Koren:2009:Computer} and the recent \textit{light  graph convolution network} (LightGCN)~\cite{Xiangnan:2020:SIGIR} as two recommendation models. For a fair comparison, we set the identical parameters of the recommendation model for all comparing negative sampling algorithms. The codes is implemented with Numpy and Pytorch, respectively. Computations were conducted on a personal computer with Windows 10 operating system, 2.1 GHz CPU,  RTX 1080Ti GPU, and 32 GB RAM.


\subsubsection{Evaluation Metrics}
To examine sampling quality, we measure the quality of the sampled instance from two perspectives: \textit{sampling bias rate} and \textit{average loss gradient magnitude}. By flipping labels of ground-truth records in the test set, we are able to obtain the false negatives (FN) that are positive labeled but unobserved during the negative sampling process. And the rest of un-interacted items are true negatives (TN). For each epoch, we record each sampled instance's label and loss gradient magnitude $\mathtt{info}(j)$, then define the unbiasedness and informativeness epoch-wisely by:
\begin{eqnarray}
	TNR &=& \frac{\#TN}{\#TN+ \#FN}, \label{Eq:TNR}\\
	INF &=& \frac{ \sum_j \mathtt{info}(j) \cdot \mathtt{sgn}(j)}{\#TN+ \#FN}, \label{Eq:INF}
\end{eqnarray}
where $\#TN$($\#FN$) is the number of sampled true (false) negatives in each training epoch. Eq.~\eqref{Eq:TNR} evaluates the proportion of sampled true negatives in each training epoch, i.e., true negative rate (TNR). $\mathtt{sgn}(j)$ is the indicator function: $\mathtt{sgn}(j)=1$ if the sampled item's label is TN; Otherwise, $\mathtt{sgn}(j)=-1$ as a penalty for sampling the FN instance. The \textit{informativeness} (INF) defined Eq.~\eqref{Eq:INF} can be interpreted as the average gradient magnitude with respect to selected training triple $(u,i,j)$ in each training epoch. To evaluate recommendation performance, the widely used metrics are adopted, including P(precision), R(recall), NDCG (normalized discounted cumulative gain), to evaluate the Top-$K$ recommendation. For their common usage, we do not provide their definitions here.

\subsection{Experiment Results}
\subsubsection{Recommendation Performance}
Implementation details: (a) MF~\cite{Xiangnan:2016SIGIR}: embedding dimension $d=32$, learning rate $\alpha = 0.01$, regulation constant $\lambda = 0.01$ and training epoch $T=100$, batch size $b=1$. (b) Light GCN~\cite{Xiangnan:2020:SIGIR}:  embedding dimension $d=32$, the initial learning rate $\alpha = 0.01$ and decays every 20 epochs with the decay rate=0.1, regulation constant $reg = 10^{-5}$, number of LightGCN layers $l = 1$, training epoch $T=100$,  batch size $b=128$ for MovieLens-100K and Yahoo!-R3 datasets, $b=1024$ for MovieLens-1M. 

Table~\ref{Table:Recommendation} compares the recommendation performance for the negative sampling algorithms, where the boldface and underline are used to indicate the best and the second best in each comparing group. The proposed \textsf{BNS} algorithm achieves the best performance in almost all cases (except two second best) of the two recommendation models, three testing datasets and three performance metrics. The results validate that our algorithm can sample high-quality negatives measured from both informativeness and unbiasedness. It is noted that the LightGCN recommendation model outperforms the MF one in general, which should thank its use of graph structure and powerful neural model for representation learning. We have run our \textsf{BNS} for 10 times, the standard deviations for each evaluation metric are consistently less than 0.002.

\begin{table*}[h!]
	\centering
	\caption{Comparison of recommendation performance on the three datasets.}\label{Table:Recommendation}
	\resizebox{1\textwidth}{!}{
		\begin{tabular}{lclccccccccccc}
			\toprule[1.2pt]
			\multirow{2}*{\textbf{Dataset}} & \multirow{2}*{\textbf{CF Model}} & \multirow{2}*{\textbf{ Method}} & \multicolumn{3}{c}{Top-5} &~& \multicolumn{3}{c}{Top-10}&~&\multicolumn{3}{c}{Top-20}\\ \cline{4-6} \cline{8-10} \cline{12-14}
			
			~ & ~ & ~ & Precision& Recall& NDCG& ~ &Precision& Recall& NDCG& ~ &Precision& Recall& NDCG \\ \hline
			
			\multirow{12}*{\textbf{100K}} & \multirow{6}*{\textbf{MF}} & RNS & 0.3900   &0.1301	&0.4143	&~&0.3363	&0.2164	&0.3967& ~&0.2724&0.3298&0.3962 \\
			~ & ~ & PNS  &0.2647	&0.0864	&0.2694	&~&0.2329	&0.1475	&0.2637& ~&0.1949&0.2374&0.2709\\
			~ & ~ & AOBPR  &0.3970	&0.1375	&0.4186&~	&0.3308	&0.2165	&0.3942& ~&0.2700&0.3369&0.3980\\
			~ & ~ & DNS  &\underline{0.4053}	&\underline{0.1414}	&\underline{0.4314} &~	&0.3348	&\underline{0.2214}	&\underline{0.4042}& ~&0.2734&\underline{0.3413}&\underline{0.4069}\\
			~ & ~ & SRNS  &0.3951	&0.1342	&0.4176&~	&\underline{0.3394}	&0.2174	&0.3998& ~&\underline{0.2747}&0.3374&0.4013 \\
			
			~ & ~ &Proposed    &\textbf{0.4205	}&\textbf{0.1467}	&\textbf{0.4558}&~	&\textbf{0.3463}	&\textbf{0.2290}	&\textbf{0.4217}& ~&\textbf{0.2762}&\textbf{0.3466}& \textbf{0.4176}\\
			\cline{2-14}

			~ & \multirow{6}*{\textbf{LightGCN}} & RNS & 0.4261&0.1453&0.4544&~&0.3571&0.2319&0.4275&~& 0.2867&0.3490&0.4248\\
			~ & ~ & PNS & 0.3527&0.1266&0.3816&~&0.3015&0.2117&0.3660& ~& 0.2461&0.3306&0.3742\\
			~ & ~ & AOBPR & 0.3911&0.1407&0.4200&~&0.3315&0.2276&0.4007&~&0.2680&0.3505&0.4064\\
			~ & ~ & DNS & \underline{0.4278}& \underline{0.1475}&\underline{0.4590}&~&\underline{0.3612}&\underline{0.2336}&\underline{0.4331}& ~&\textbf{0.2917}&\underline{0.3595}&\underline{0.4335}\\
			~ & ~ & SRNS & 0.4195&0.1440&0.4509&~&0.3564&0.2333&0.4275& ~&0.2834&0.3520&0.4244\\
			
			~ & ~ & Proposed & \textbf{0.4318}&\textbf{0.1518}&\textbf{0.4640}&~& \textbf{0.3671}&\textbf{0.2410}&\textbf{0.4368}& ~&\underline{0.2875}& \textbf{0.3608}&\textbf{0.4383}\\
			\bottomrule[1.0pt]

			\multirow{12}*{\textbf{1M}} & \multirow{6}*{\textbf{MF}} & RNS & 0.3843    &0.0855	&0.4027	&~&0.3353	&0.1430	&0.3737& ~&0.2798&0.2244&0.3572 \\
			~ & ~ & PNS  &0.3461	& 0.0753&0.3634	&~&0.3004	&0.1250	&0.3356& ~&0.2502&0.1979&0.3192\\
			~ & ~ & AOBPR & 0.3946&0.0954&0.4135&~&0.3416&0.1549&0.3837& ~&0.2857&0.2442&0.3714\\
			~ & ~ & DNS  &\underline{0.4066}	&\underline{0.0991}	&\underline{0.4272}&~	&\underline{0.3521}	&\underline{0.1620}	&\underline{0.3965}& ~&\underline{0.2945}&\underline{0.2537}&\underline{0.3838} \\
			~ & ~ & SRNS  &0.3955	&0.0934	&0.4225&~	&0.3408&0.1609	&0.4042& ~&0.2779&0.2431&0.3974\\
			~ & ~ & Proposed  &\textbf{0.4207}	&\textbf{0.1062}	&\textbf{0.4324}&~	&\textbf{0.3518}	&\textbf{0.1703}	&\textbf{0.4191}& ~&\textbf{0.3045}&\textbf{0.2614}&\textbf{0.4002}\\ \cline{2-14}

			~ & \multirow{6}*{\textbf{LightGCN}} & RNS &0.4095&0.0953&0.4305&~&0.3512&0.1547&0.3985& ~&0.2915&0.2405&0.3781 \\
			~ & ~ & PNS  &0.3658	& 0.0907&0.3855	&~&0.3152	&0.1486	&0.3564& ~&0.2608&0.2314&0.3440\\
			
			~ & ~ & AOBPR  &0.4073	&\underline{ 0.0997}	&0.4286&~	&0.3535	&\textbf{0.1626}&0.3982& ~&0.2949&\underline{0.2536}&\underline{0.3849}\\
			~ & ~ & DNS &\underline{0.4130}  &0.0972&\underline{0.4342} &~&\underline{0.3552}&0.1577&\underline{0.4002}& ~&\underline{0.2958}&0.2468&0.3840\\
			~ & ~ & SRNS & 0.4026&0.0973&0.4239&~&0.3515&0.1526&0.3953&~&0.2922&0.2524&0.3815\\
			~ & ~ & Proposed & \textbf{0.4228}&\textbf{0.1087}&\textbf{0.4438}&~&\textbf{0.3639}&\underline{0.1612}&\textbf{0.4088}& ~&\textbf{0.3025}&\textbf{0.2527}&\textbf{0.3917}\\
			\bottomrule[1.0pt]

			\multirow{12}*{\textbf{Yahoo}} & \multirow{6}*{\textbf{MF}} &RNS & 0.1196   &0.0875&0.1326	&~&0.0935	&0.1367	&0.1401 &~&0.0695&0.2015&0.1665 \\
			~ & ~ & PNS & 0.1186&0.0876&0.1301&~&0.0927&0.1360&0.1378&~&0.0688&0.2011&0.1644\\
			~ & ~ & AOBPR & 0.1012&0.0741&0.1115&~&0.0798&0.1165&0.1184&~&0.0607&0.1778&0.1443\\
			~ & ~ & DNS &\underline{0.1251} &\underline{0.0917}&\underline{0.1390}&~&\underline{0.0957}&\underline{0.1399}&\underline{0.1449}&~&\underline{0.0697}&0.2020&\underline{0.1697}\\
			~ & ~ & SRNS &0.1141&0.0855&0.1285&~&0.0904&0.1358&0.1383&~& 0.0678&\underline{0.2025}&0.1655\\
			~ & ~ & Proposed &\textbf{ 0.1303}&\textbf{0.0975}&\textbf{0.1470}&~&\textbf{0.1002}&\textbf{0.1485}&\textbf{0.1542}&~& \textbf{0.0711}&\textbf{0.2094}&\textbf{0.1783}\\ \cline{2-14}

			~ & \multirow{6}*{\textbf{LightGCN}} & RNS & 0.1479&0.1101&0.1693&~&0.1126&0.1669&0.1760&~& 0.0814&0.2389&0.2047\\
			~ & ~ & PNS & 0.1076&0.0797&0.1214&~&0.0809&0.1185&0.1254&~&0.0590&0.1708&~0.1464\\
			~ & ~ & AOBPR & 0.1462&0.1120&0.1635&~&	0.1048&0.1552&0.1612&~& 0.0763&0.2229&0.1886\\
			~ & ~ & DNS & \underline{0.1530}&\underline{0.1137}&\underline{0.1743}&~&\underline{0.1148}&\underline{0.1697}&\underline{0.1800} &~&\underline{0.0829} &\underline{0.2433}&\underline{0.2089}\\
			~ & ~ & SRNS & 0.1457&0.1092&0.1668&~&0.1121&0.1636&0.1735&~& 0.0799&0.2352&0.2017\\
			~ & ~ & Proposed &\textbf{ 0.1550}&\textbf{0.1157}&\textbf{0.1768}&~&\textbf{0.1169}&\textbf{0.1729}&\textbf{0.1827}&~&\textbf{0.0837}&\textbf{0.2459}&\textbf{0.2117}\\
			\cline{1-14}
			
			\bottomrule[1.2pt]
			
		\end{tabular}
	}
\end{table*}

\par
Take a close look on the results. It is interested to find that among the two static negative sampling algorithms, the \textsf{RNS} generally outperforms the \textsf{PNS}. This suggests that the  popularity-based sampling distribution favoring popular items may actually introduce more biases in negative sampling. Among the three hard negative sampling algorithms, viz., \textsf{AOBPR}, \textsf{DNS} and \textsf{SRNS}, it is noted that the \textsf{DNS} often outperforms the other two.  The \textsf{AOBPR} prioritizes those global higher ranked items, While the \textsf{DNS} first randomly selects a few negatives, among which favors those local relatively higher ranked items. Since the \textsf{DNS} balances between informativeness and unbiasedness to some extent, it can achieve the second best in many cases. The \textsf{SRNS} exploits the empirical observation that a negative with high-variance of its predicted scores could be a true negative. Although it is an interesting approach to identify true negative, the linear average operation of \textsf{SRNS} for selecting high quality instances may weaken its effectiveness.

\subsubsection{Negative Sampling Quality}
\begin{figure*}[h!]
	\centering
	\includegraphics[width=\textwidth]{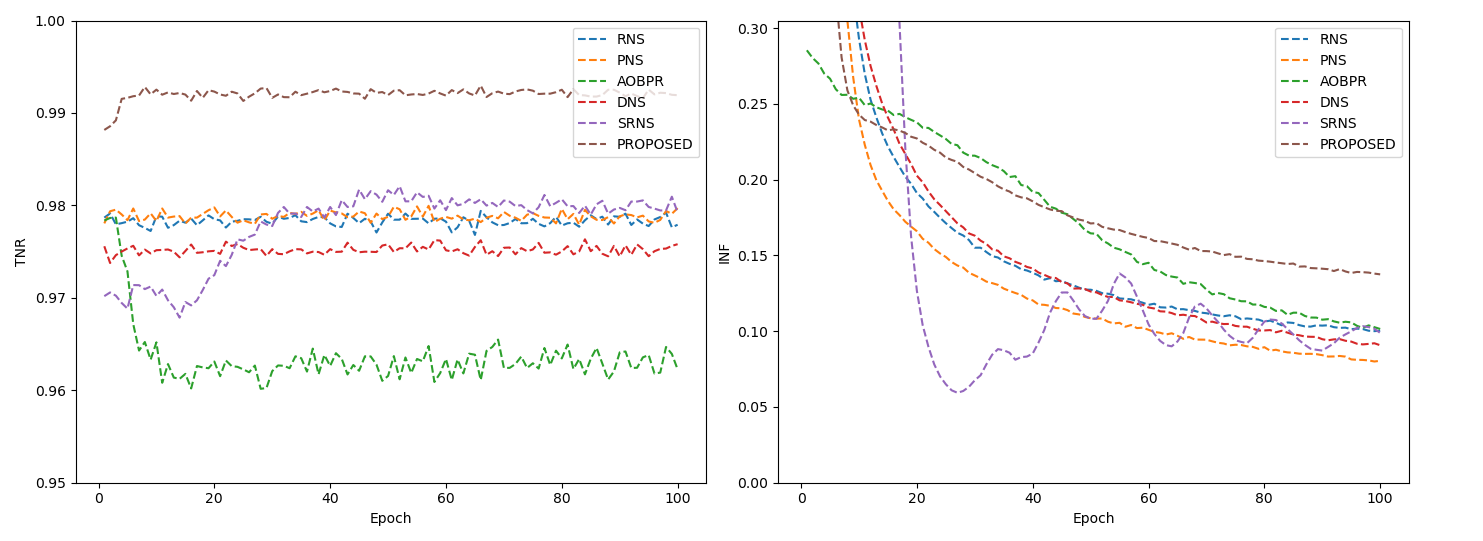}
	\caption{The quality of selected negatives in each training epoch on the MovieLens-100K dataset.}
	\label{Fig:NegQua}
\end{figure*}
Implementation details: In this setion, we test two sampling criteria in terms of sampling bias and sampling quality: (i) the \textit{posterior probability criterion} given by equation Eq.~\eqref{Eq:unbias}. (ii) \textit{Bayesian sampling criterion} given by Eq.~\eqref{Eq:NegativeSam}. The former aims to select true negative instances, while the latter aims to sample high-quality negative instances. The reliability of the $\mathtt{unbias}(l)$ estimation is crucial for negative sampling. We first examine whether \textit{posterior probability criterion} is capable of selecting true negative instances, and then examine whether \textit{Bayesian sampling criterion} is capable of selecting high quality negatives. 
 
The \textit{posterior probability criterion }is achieved by selecting the negative instance with the largest $\mathtt{unbias}(\cdot)$-value from the candidate set $\mathcal{M}_u$ :
\begin{eqnarray} \label{Eq:PosteriorSam}
	j   &=&   \mathop{\arg\max}\limits_{l \in \mathcal{M}_u} \mathtt{unbias}(l) 
\end{eqnarray}
$\mathcal{I}_u^-$ is a small candidate set containing randomly selected  negative instances from $\mathcal{I}_u^-$. The sampling quality of different sampling methods is exhibited in Fig~\ref{Fig:NegQua}.

The \textit{Bayesian sampling criterion} is achieved by selecting the negative instance with the smallest  $R(\cdot|i) $-value from the candidate set $\mathcal{M}_u$ :
\begin{eqnarray} \label{Eq:NegativeSam1}
	j  = \mathop{\arg\min}\limits_{l \in\mathcal{M}_u}~ \mathtt{info}(l)\cdot [1-(1+\lambda)\mathtt{unbias}(l)]
\end{eqnarray}
For above sampling methods, the size of $\mathcal{M}_u$ is fixed as 5.

\par
(i) Sampling bias: \textit{Fixed distribution sampling} (\textsf{RNS} and \textsf{PNS}) achieves relatively moderate performance. Their TNRs fluctuate around the probability of a random sample being a true negative. \textit{Hard negative sampling} (\textsf{AOBPR} and \textsf{DNS}) has the worst performance. They adopt a greedy strategy to emphasize higher ranked negatives, also bringing higher risk of sampling false negatives per our discussion in Section~\ref{Sec:Dis}. The \textsf{SRNS} uses simple prior statistic information of variance of predicted scores, which limits the potentials of negative classification, because this prior variance may result in the sampling distribution to be overly concentrated. The proposed \textit{Bayesian negative sampling} (\textsf{BNS}) achieves the best performance for its TNR closer to 1, owing to our Bayesian negative classification.

(ii) Sampling quality: The INF decreases with the increase of training epoches. This is because the trained recommendation model can rank the false negatives potentially interested by users higher than true negatives (cf. Fig.~\ref{Fig:DistributionStatistics}). Our \textsf{BNS} achieves the best performance after enough training epoches.  The \textit{hard negative sampling} (\textsf{AOBPR} and \textsf{DNS}) suffer more penalties due to its highest sampling bias. The \textsf{SRNS} adopts a linearly weighted average to combine informativeness and variance, which may not guarantee sampling unbiased and informative instances.

\subsection{Study of \textsf{BNS}}
We perform sensitive analysis to get deep insights on \textsf{BNS}.

\subsubsection{Hyper-parameter selection}\label{Appendix:Selection}
\begin{figure*}[h]
	\centering
	\includegraphics[width=\textwidth]{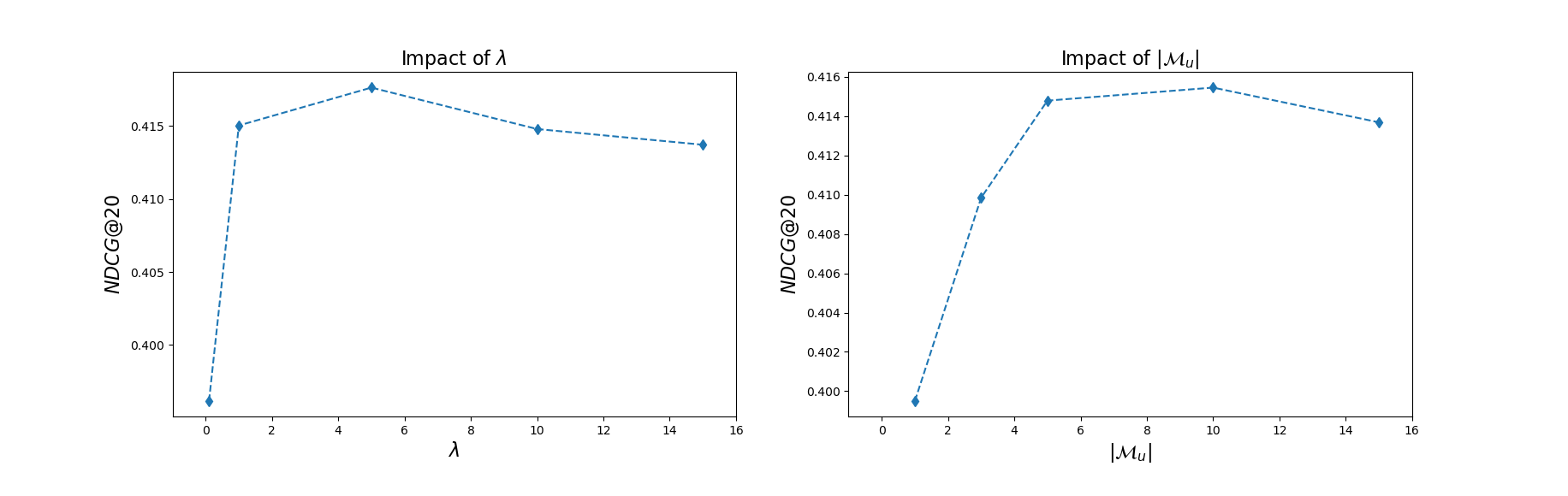}
	\caption{The impact of $\lambda$ and $|\mathcal{M}_u|$ w.r.t NDCG metric.}
	\label{Fig:highperparameter}
\end{figure*}

We first fix the size of candidate set $\mathcal{M}_u$ as 5 to investigate how $\lambda$ affects the performance. In particular, we search $\lambda$ in the range of \{0.1, 1, 5, 10, 15\}. A larger value of $\lambda$ means that putting more emphasis on the ranking gain from true negatives, and conversely more attention is paid to avoiding the risk of sampling false negatives if $\lambda$ is small. We can observe from Fig.~\ref{Fig:highperparameter} that, when the value of $\lambda$ increases from 0.1 to 1, $NDCG@20$ improves significantly, and achieves its maximum value when $ \lambda= 5$. This does not mean that the optimal a-value is fixed during the model training.  

Then we fix the optimal $\lambda$ as 5 to study the impact of size of $\mathcal{M}_u$, and search $|\mathcal{M}_u|$ in the range of \{1, 3, 5, 10, 15\}. Note that when $|\mathcal{M}_u|=1$, the proposed \textsf{BNS} reduced to the classical random negative sampling (\textsf{RNS}).  When $|\mathcal{M}_u| > 1$, the  Bayesian sampling criterion begins to play its role for selecting hard negatives. When $|\mathcal{M}_u| = |\mathcal{I}_u^-|$, the sampler $h$ of Eq~\eqref{Eq:NegativeSam} is optimal sampler $h^*$ given by Eq~\eqref{Eq:OptimalSam}. We can observe from Fig.~\ref{Fig:highperparameter} that $NDCG@20$ achieves its maximum value when $ |\mathcal{M}_u|= 5~or 10$. To reduce the time complexity, we set $|\mathcal{M}_u|= 5$.

The value of $NDCG@20$ decreases  $|\mathcal{M}_u| > 10 $. This experimental result is not as expected, since the larger value of $|\mathcal{M}_u|$, the easier it is to select the optimal negative. We believe that it is caused by the unreliable prior probability $P_{fn}(l)$ estimation. The unreliable prior information further results in the biased classification results denoted by $F(\hat{x}_l)$, leading to further deviation of negative signal $\mathtt{unbias}(\cdot)$. An excessive $|\mathcal{M}_u|$ amplifies the adverse effect of negative signal bias, leads to performance degradation. More discussions can be found in  Sec~\ref{Appendix:Asymptotic}.

\subsubsection{Sensitivity analysis}\label{Appendix:Sensitive}
We further study the sensitivity of \textsf{BNS} to $\lambda$, sample information $\hat{x}_l$, and prior information $P_{fn}(l)$. 
\begin{itemize}
	\item[-]\textsf{BNS}-1: warm-start of $\lambda$. We set  $\lambda = max(10-\alpha*epoch,2)$ to linearly decrease the value of $\lambda$ as epoch number increases. That is, a larger $\lambda$ to emphasis the ranking gain for sampling true negative in the initial stage, and a smaller $\lambda$ to emphasis the loss for sampling false negatives in the later stages, where $\alpha$ is selected as 0.1.
	\item[-]\textsf{BNS}-2: warm-start of sample information $\hat{x}_l$. We first adopt \textsf{RNS} to train a recommendation model for some epochs to learn a more reliable sample information $F(\hat{x}_l)$, and then resume the training by replacing it with our \textsf{BNS}. 
	\item[-]\textsf{BNS}-3: non-information prior distribution. For this case, \textsf{BNS} reduces to \textsf{DNS} that use only sample information $\hat{x}_l$ for negative sampling. For a single randomized trial, the probability of any item $l$ been interacted is 1/1682, i.e., we set $ P_{fn}(l) =1/1682$ indiscriminately for any negative $l$, where 1682 is the total number of items.
	\item[-]\textsf{BNS}-4: occupation information enhanced prior distribution. On the basis of Eq~\eqref{Eq:Prior}, We added a adjustment factor to improve the estimation of $P_{fn}(l)$:
	\[P_{fn} (l) =\frac{pop_l}{N} \cdot (1+\triangle o_{ul}), \]
	where $\triangle o_{ul} = \frac{o_{ul} - \bar{o}_l }{ \max {o}_{l}}$ indicating the deviation of the number of times of $l$ has been preferred by $u$'s occupation from mean value. $o_{ul}$ is the number of interactions of groups with same occupation as $u$ on item $l$, $\bar{o}_l$ is the mean number of times of item $l$ being interacted with by each occupation.
\end{itemize}
The experimental results are exhibited in Table~\ref{Exp:study}. From which we have the following findings:
\begin{table*}[h]
	\centering
	\caption{Study of BNS .}\label{Exp:study}
	\resizebox{1\textwidth}{!}{
		\begin{tabular}{lclccccccccccc}
			\toprule[1.2pt]
			\multirow{2}*{\textbf{Dataset}} & \multirow{2}*{\textbf{CF Model}} & \multirow{2}*{\textbf{ Method}} & \multicolumn{3}{c}{Top-5} &~& \multicolumn{3}{c}{Top-10}&~&\multicolumn{3}{c}{Top-20}\\ \cline{4-6} \cline{8-10} \cline{12-14}
			
			~ & ~ & ~ & Precision& Recall& NDCG& ~ &Precision& Recall& NDCG& ~ &Precision& Recall& NDCG \\ \hline
			
			\multirow{6}*{\textbf{100K}} & \multirow{6}*{\textbf{MF}} & RNS & 0.3900   &0.1301	&0.4143	&~&0.3363	&0.2164	&0.3967& ~&0.2724&0.3298&0.3962 \\
			~ & ~ & BNS    &0.4205	&0.1467	&0.4558&~	&0.3463	&0.2290	&0.4217& ~&0.2762&0.3466& 0.4176\\
			~ & ~ & BNS-1  &0.4237	&0.1471	&0.4551	&~&0.3495	&0.2305	&0.4238& ~&0.2762&0.3495&0.4197\\
			~ & ~ & BNS-2  &0.4148	&0.1456	&0.4449&~	&0.3411	&0.2245 &0.4132& ~&0.2738&0.3434&0.4125\\
			~ & ~ & BNS-3  &0.4048	&0.1392	&0.4266&~	&0.3423	&0.2282 &0.4043& ~&0.2720&0.3406&0.4030\\
			~ & ~ & BNS-4  &0.4262	&0.1478	&0.4566&~	&0.3486	&0.2305 &0.4235& ~&0.2792&0.3520&0.4216\\
			\cline{1-14}
			
			\bottomrule[1.2pt]
			
		\end{tabular}
	}
\end{table*}

\textbf{Sensitivity of $\lambda$ to \textsf{BNS}}: The warm start of $\lambda$ achieved better performance (\textsf{BNS}-1). As the degree of trade-off between sampling risk and gain, a larger value of $\lambda$ means that we place more emphasis on the ranking gain from TN than rather than the risk from FN. The results show that hard negative instances are important for model learning, which is consistent with the findings of existing studies~\cite{Ding:2020:NIPS,Park:2019:WWW}. We recommend a warm-start strategy of $\lambda$: larger value to emphasis the hard negatives in the initial stage, and small $\lambda$ to avoid sampling false negatives in later stage.

\textbf{Sensitivity of prior probability $P_{fn}(\cdot)$ to \textsf{BNS}}: The results shows that \textsf{BNS-3} achieved worse performance  in the absence of prior information compared with standard \textsf{BNS}, while \textsf{BNS-4} achieved better performance in the case of occupation-enhanced prior probability compared with standard \textsf{BNS}. The results show that \textsf{BNS} is sensitive to priori probability. The mechanism by which priori information affects negative sampling is: the unreliable prior probability $P_{fn}(\cdot)$ results in the biased classification results denoted by $F(\hat{x}_\cdot)$, leading to further deviation of negative signal $\mathtt{unbias}(\cdot)$. An excessive size of candidate set $\mathcal{M}_u$ amplifies the adverse effect of negative signal bias, leads to performance degradation. Therefore, the larger size of $\mathcal{M}_u$ is the better if the prior probability $P_{fn}(\cdot)$ is reliable, otherwise an $\mathcal{M}_u$ of moderate size should be chosen. In particular, \textsf{BNS} is equivalent to \textsf{DNS} in the case of non-information prior distribution:   \textsf{BNS} samples instances with appropriate $F(\hat{x})$-values and the \textsf{DNS} samples instances with appropriate ranking positions, while $F(\hat{x})$ and ranking position are with one-to-one mapping. By selecting the appropriate $|\mathcal{M}_u|$ and $\lambda$, \textsf{BNS-3} achieves comparable performance with \textsf{DNS}. We recommend to select the most reliable priori information for modeling  $P_{tn}(\cdot)$ or  $P_{fn}(\cdot)$.

\textbf{Sensitivity of sample information $\hat{x}_l$ to \textsf{BNS}}: The warm start of the sample information $\hat{x}$ (\textsf{BNS-2}) did not achieve better performance than standard \textsf{BNS} as expected, we believe there are three reasons: (i) random sampling at the initial training stages is difficult to sample hard negatives, making the performance of \textsf{BNS-2} degrade; (ii)  $\hat{x}$ is endogenously determined by priori probability and the sampler $h$, therefore the warm-start of $\hat{x}$ has limited impact on the final ranking performance. (iii) the way we use sample information  $F(\hat{x}_\cdot)$ is insensitive to small changes in $\hat{x}_\cdot$, thus the improved $\hat{x}$ has limited impact on improving the sampling quality. We believe that this property of \textsf{BNS} is an important manifestation of its robustness for different ranking models, as \textsf{BNS} is still able to sample high quality negative instances using prior information under the condition of the order relation does not hold (e.g., early training stage or weak ranking models).

\subsubsection{Asymptotic optimal sampler}\label{Appendix:Asymptotic}
Next, we will show the asymptotic process of proposed sampler $h$ to the optimal sampler $h^*$ given the ideal prior probability $P_{fn}(l)$. We set $P_{fn}(l) = (label(l)-0.2)^2$, that is , $P_{fn}(l)=0.64$ if $l\in fn$ otherwise $P_{fn}(l)=0.04$. The asymptotic process is achieved by gradually increasing the size of $\mathcal{M}_u$, where $\lambda$ is fixed as 5. Simulation results are presented in Table~\ref{Table:asymptoticprocess}. By increasing the size of the candidate set size, the optimal sampler $h^*$ is achieved without ranking performance degradation. This result validates our analysis in Sec~\ref{Appendix:Selection}. Equipped with certain degree of a priori information, \textsf{BNS} achieved considerable performance even for simple dot-product based representation learning methods, demonstrating the great potential of negative sampling studies. The performance of optimal sampler (i.e., $|\mathcal{M}_u| = |\mathcal{I}_u^-|$)  is an empirical upper bound for dot product-based model. Due to the existence of low-rank constraint thus limited expressiveness of matrix factorization, the recommendation performance cannot reach 1.

\begin{table*}[h]
	\centering
	\caption{ The asymptotic process to the optimal sampler $h^*$.}\label{Table:asymptoticprocess}
	\resizebox{1\textwidth}{!}{
		\begin{tabular}{lclccccccccccc}
			\toprule[1.2pt]
			\multirow{2}*{\textbf{Dataset}} & \multirow{2}*{\textbf{CF Model}} & \multirow{2}*{\textbf{\textsf{BNS} Size}} & \multicolumn{3}{c}{Top-5} &~& \multicolumn{3}{c}{Top-10}&~&\multicolumn{3}{c}{Top-20}\\ \cline{4-6} \cline{8-10} \cline{12-14}
			
			~ & ~ & ~ & Precision& Recall& NDCG& ~ &Precision& Recall& NDCG& ~ &Precision& Recall& NDCG \\ \hline
			
			\multirow{9}*{\textbf{100K}} & \multirow{9}*{\textbf{MF}} & $|\mathcal{M}_u| = 1$ & 0.3900   &0.1301	&0.4143	&~&0.3363	&0.2164	&0.3967& ~&0.2724&0.3298&0.3962 \\
			~ & ~ & $|\mathcal{M}_u| = 3$  &0.4909	&0.1567&0.5211&~	&0.4220	&0.2565	&0.4942& ~&0.3366&0.3872&0.4856\\
			~ & ~ & $|\mathcal{M}_u| = 5$ &0.5109	&0.1612	&0.5422	&~&0.4329	&0.2602	&0.5092& ~&0.3456&0.3925&0.4992\\
			~ & ~ & $|\mathcal{M}_u| = 10$  &0.5351	&0.1696	&0.5685&~	&0.4589	&0.2722 &0.5365& ~&0.3663&0.4081&0.5245\\
			~ & ~ & $|\mathcal{M}_u| = 20$  &0.5760	&0.1828	&0.6070&~	&0.4885	&0.2875 &0.5695& ~&0.3830&0.4196&0.5498\\
			~ & ~ & $|\mathcal{M}_u| = 50$  &0.6239	&0.1989	&0.6599&~	&0.5252	&0.3049 &0.6146& ~&0.4031&0.4312&0.5843\\
			~ & ~ & $|\mathcal{M}_u| = 100$   &0.6509	&0.2104	&0.6898&~	&0.5382	&0.3125 &0.6346& ~&0.4053&0.4321&0.5971\\
			~ & ~ & $|\mathcal{M}_u|= 500$  &0.6661	&0.2183	&0.7128&~	&0.5412	&0.3131 &0.6487& ~&0.4041&0.4300& 0.6076\\
			~ & ~ & $|\mathcal{M}_u|= |\mathcal{I}_u^-|$  &0.6674	&0.2184	&0.7133&~	&0.5429	&0.3140 &0.6495& ~&0.4041& 0.4292&0.6073\\
			\cline{1-14}
			
			\bottomrule[1.2pt]
			
		\end{tabular}
	}
\end{table*}

%

\subsection{In-depth Analysis}
The baselines contain the results of two illuminating ablation experiments: (i) \textsf{PNS} that only uses prior information (item popularity), and (ii) \textsf{DNS}, \textsf{AOBPR} and \textsf{SRNS} that only use sample information $\hat{x}_l$. In particular, \textsf{BNS} reduces to \textsf{DNS} in the case of an non-information prior distribution. \textsf{BNS} samples instances with appropriate $F(\hat{x})$-values and the \textsf{DNS} samples instances with appropriate ranking positions, while $F(\hat{x})$ and ranking position are with one-to-one mapping, so they achieved comparable performance. The shortcomings of the above two paradigms are obvious: the former can incorporate domain knowledge but independent of model status, resulting in a static sampling distribution; the latter exploits sample information but ignore priori knowledge, which will degrade performance especially in the scenarios of rich side information. The proposed \textsf{BNS} combines the priori information and sample information from the Bayesian perspective, and define the negative signal $\mathtt{unbias}(l)$ in a posterior probability sense. In turn, the optimal sampling rule can be given. We believe that the performance improvement of \textsf{BNS} stems from three aspects: (i) the prior information $P_{tn}(l)$ or $P_{fn}(l)$ , and (ii) the use of sample information $\hat{x}_l$ that avoids negative sampling to conflict with classification results of the ranking model. (iii) \textsf{BNS} is theoretically optimal sampler that minimizes the empirical sampling risk.

\section{Related Work}
Pairwise learning and pairwise loss have been widely applied in recommendation systems~\cite{Steffen:2009:UAI,Weike:2013:IJCAI,Yu:2018:CIKM,Steffen:2014:WSDM}. Pairwise comparisons of positive instances and negative instances are first constructed to train a recommendation model. How to select negatives for pairwise comparisons, i.e., negative sampling, is a key to model training~\cite{Steffen:2014:WSDM,Zhang:2013:SIGIR,Ding:2020:NIPS}. We review the related work of negative sampling for recommendation from two categories according to whether the sampling policy is fixed during the model training process.

\par
The first category is the \textit{static negative sampling}. This kind of methods adopt a fixed sampling distribution for negative sampling during the whole training process. The most widely used is the random negative sampling (RNS)~\cite{Steffen:2009:UAI,Xuejiao:2020:ASC,Yu:2018:CIKM,Xiangnan:2020:SIGIR}, which uniformly samples negatives from un-labeled instances. Some have proposed to set sampling probability of a negative instance according to its popularity (interaction frequency), as so-called popularity-biased negative sampling(PNS)~\cite{Mikolov:2013:NIPS,Chen:2017:KDD,
	Tang:2015:WWW,Mihajlo:2015:SIGIR,Caselles:2018:Recsys}. Among them, the most widely used sampling distribution is $p(j) \propto r_j^{0.75}$, where $r_j$ is the interaction frequency of an item in the training dataset.

\par
The second category is the \textit{hard negative sampling} that adopts an adaptive sampling distribution targeting on sampling hard negative instances. The so-called hard negatives refer to those unlabeled instances that are similar to those positive instances in the embedding space~\cite{Rao:2016:CIKM,Zhang:2019:ICDE,Sun:2018:IJCAI}. Many hard negative sampling strategies have been proposed for personalized recommendation~\cite{Steffen:2014:WSDM,Zhang:2013:SIGIR,Ding:2020:NIPS,Park:2019:WWW,Huang:2021:KDD,Ding:2019:IJCAI}. For example, Zhang et al~\cite{Zhang:2013:SIGIR} and Steffen et al~\cite{Steffen:2014:WSDM} propose to oversample higher scored thus higher ranked negatives that are argued to be more similar to positive instances. Some have proposed to exploit graph-based information for boosting negative sampling~\cite{Wang:2020:WWW, Wang:2021:CIKM, Li:2014:WWW,Chen:2019:WWW,Wang:2021:TKDE,Ying:2018:KDD}. For example, Wang et al~\cite{Wang:2020:WWW} and Wang et al~\cite{Wang:2021:CIKM} propose to leverage the types of relations on a knowledge graph to filter hard negatives. Another approach is to use the random walk on a graph for selecting hard negatives that are structurally similar to positive instances~\cite{Li:2014:WWW,Chen:2019:WWW,Wang:2021:TKDE,Ying:2018:KDD}.

\par
Additional or prior information can be exploited for identifying and sampling hard negatives,  Such side information are intuitive for distilling negative signals, such as users' connections in social networks~\cite{Zhao:2014:CIKM,Wang:2016:CIKM},  geographical locations of users~\cite{Yuan:2016:IJCAI,Liu:2019:IJCAI}, and additional interaction data such as viewed but non-clicked~\cite{Jingtao:2019:IJCAI, Jingtao:2018:WWW}. Beside only sampling negatives from unlabeled instances, a novel class of methods is to generate a kind of virtual hard negatives from multiple negative instances or by using some generative method. For example, Huang et al.~\cite{Huang:2021:KDD} propose to synthesize virtual hard negatives by hop mixing embeddings. Jun et al.~\cite{Jun:2017:SIGIR} and Park et al.~\cite{Park:2019:WWW} design generative adversarial neural networks for generating virtual hard negatives.
\section{Conclusion}\label{Sec:Conclusion}
This paper has provided a comprehensive analysis on negative sampling for recommendation. On the basis of order relation analysis of negatives' scores, we derive the class conditional density of true negatives and that of false negatives, and provide an affirmative answer from a Bayesian viewpoint to distinguish true negatives from false negatives. Then, according to the  asymptotic property of the empirical distribution function, we defined a model-agnostic posterior probability estimate of an instance being true negative as a quantitative negative signal measure. Lastly, we propose a Bayesian sampling rule to sample high-quality negative instances. It is the theoretically optimal sampling rule that minimizes the empirical sampling risk. Experiment studies have validated our arguments and findings.

\par
The limitations of \textsf{BNS} are: (i) We adopt a simplistic approach for modeling prior probability, yet it is critical to the Bayesian-based methods. (ii) Due to the collaborative filtering mechanism, it is difficult to obtain the analytical solution of sampling loss $\triangle \mathcal{L}(l|i)$ from single update of a train triple. Eq~\eqref{Eq:rankinggain} is just an approximation of sampling loss, which has much room for improvement. We note that negative sampling indeed faces an exploration-and-exploitation trade-off: Exploration suggests to prioritize those higher ranked informational instances been classified as false negative (positive labeled) by a ranking model; Exploitation indicates to favor those lower ranked unbiased instances been classified as true negative by the same ranking model.  Future work can go further to generalize \textsf{BNS} to contrastive-based learning methods and investigate the optimal trade-off between such exploration and exploitation.


\normalem
\bibliographystyle{IEEEtran}
\bibliography{ref}
\end{document}